\documentclass[prapplied,%
 reprint,
 superscriptaddress,
 amsmath,amssymb,
 aps,
]{revtex4-2}

\usepackage{graphicx}
\usepackage{dcolumn}
\usepackage{bm}
\usepackage{hyperref}

\usepackage{graphicx}
\usepackage{dcolumn}
\usepackage{bm}
\usepackage{bbm}
\usepackage{soul}
\usepackage{enumitem}
\usepackage{makecell}

\usepackage{braket}
\usepackage{graphicx}
\usepackage{amsmath,verbatim,latexsym,amssymb,indentfirst,mathrsfs,mathtools,amsthm,bbm,bm,hyperref,url}

\usepackage{cancel}
\usepackage{bbold}
\usepackage[dvipsnames]{xcolor}
\usepackage[normalem]{ulem}
\usepackage{physics}

\begin{document}

\preprint{APS/123-QED}

\title{Anomalous Microwave Response in YBCO Resonators beyond the Two-Level-System Model}

\author{Kaiwen Zheng}
\affiliation{Department of Physics, Washington University, Saint Louis, MO, USA, 63130.}

\author{Nathan J. Johnson}
\affiliation{Department of Physics, Washington University, Saint Louis, MO, USA, 63130.}

\author{Nathan T. Thobaben}
\affiliation{Department of Electrical Engineering, Saint Louis University, Saint Louis, MO, USA, 63103.}

\author{Sidharth Duthaluru}
\affiliation{Department of Physics, Washington University, Saint Louis, MO, USA, 63130.}

\author{Haochen Shen}
\affiliation{Department of Physics, Washington University, Saint Louis, MO, USA, 63130.}
\author{Denae T. Cherry}
\affiliation{Department of Electrical Engineering, Saint Louis University, Saint Louis, MO, USA, 63103.}

\author{David S. Wisbey}
\affiliation{Department of Electrical Engineering, Saint Louis University, Saint Louis, MO, USA, 63103.}
\author{Kater W. Murch}%
 \email{katermurch@berkeley.edu}
     \affiliation{Department of Physics, Washington University, Saint Louis, MO, USA, 63130.}
    \affiliation{Department of Electrical Engineering and Computer Science, University of California Berkeley, Berkeley, CA, USA, 94720.}
    \affiliation{Department of Physics, University of California Berkeley, Berkeley, CA, USA, 94720.}

\date{\today}

\begin{abstract}

We report the microwave response of coplanar-waveguide (CPW) resonators fabricated from $\mathrm{YBa_2Cu_3O_{7-\delta}}$ (YBCO) thin films over temperatures from approximately $70~\mathrm{mK}$ to $40~\mathrm{K}$. The resonators exhibit internal quality factors $Q_\mathrm{i}$ in the range of $4\times10^3$ to $10^4$ at 70 mK, which increase to a maximum of approximately $8\times10^3$ to $1.2\times10^4$ near $6~\mathrm{K}$. At low temperatures, both $Q_\mathrm{i}$ and the fractional shift of the resonance frequency $\Delta f_\mathrm{r}/f_\mathrm{r}$ increases with temperature, qualitatively resembling behavior commonly associated with two-level-system (TLS) defects. However, neither response saturates on the temperature scale set by the resonator frequency, and the loss exhibits no observable microwave-power dependence. We show that low-temperature frequency upturn may be better described by an additional paramagnetic response associated with defect-induced local moments or Andreev bound states, while the low-temperature loss follows an approximately logarithmic temperature dependence whose microscopic origin remains unresolved. These measurements establish the millikelvin performance of patterned YBCO resonators and show that their low-temperature response cannot be understood within the conventional TLS framework alone.

\end{abstract}

\maketitle
\section{Introduction}

On-chip superconducting microwave resonators are widely used as particle detectors, readout elements for solid-state qubits, and sensitive probes of condensed-matter systems~\cite{gaoThesis, cQED_review, amir_Nb_NiFe, KC_MoTe, oliver_NbSe2, uriVool_BSCCO, triLayerGraphene}. Their resonance frequency and internal quality factor can be quantitatively related to the reactive and dissipative response of the constituent materials~\cite{gaoThesis, McRae_resonator, KC_method, KC_method2}. In resonators fabricated from conventional superconductors, low-temperature frequency shifts and loss are commonly interpreted in terms of two-level-system (TLS) defects in interface dielectric and thermally generated quasiparticles~\cite{Muller_2019, McRae_resonator, deLeon_Ta_resonator}. This framework has been extensively developed and has helped guide the optimization of superconducting resonator fabrication~\cite{deLeon_Ta_resonator}.

The same interpretation may not apply directly when the resonator is fabricated from an unconventional superconductor. Nodal quasiparticles~\cite{Hirschfeld}, impurity-induced local moments~\cite{Ames_review, Cuprate_Li_Zn_NMR}, surface Andreev bound states~\cite{ABS_theory, anlage_ABS}, and other intrinsic responses of the superconducting film can contribute to both the reactive and dissipative microwave response. These contributions may produce temperature dependences that superficially resemble conventional TLS behavior, making it important to distinguish dielectric loss from electrodynamic effects intrinsic to the superconducting film.

Among unconventional superconductors, YBCO has been studied extensively at microwave frequencies, both as a probe of the superconducting pairing symmetry and as a low-loss conductor for operation at elevated cryogenic temperatures~\cite{badass_YBCO_microwave,Hirschfeld}. More recently, its high critical temperature and resilience to magnetic fields have motivated interest in YBCO for hybrid quantum systems, spin spectroscopy, and other cryogenic microwave devices~\cite{YBCO_EPR, YBCO_EPR2, german_YBCO_HIM, sidharth}. These applications require an understanding not only of the intrinsic surface impedance of the material, but also of the loss and dispersion of lithographically patterned planar circuits under their intended operating conditions. 

In this work, we characterize CPW resonators fabricated from YBCO thin films over temperatures ranging from 70~mK to 40~K. 
At higher temperatures, $\Delta f_\mathrm{r}/f_\mathrm{r}$ and $Q_\mathrm{i}$ are well described by the expected temperature dependence of the London penetration depth and quasiparticle scattering rate. At low temperatures, however, the response is anomalous. Below approximately 6~K, both $\Delta f_\mathrm{r}/f_\mathrm{r}$ and $Q_\mathrm{i}$ increase with temperature, qualitatively resembling signatures commonly attributed to TLS defects. We show, however, a conventional TLS interpretation cannot account for the low-temperature behavior: $Q_\mathrm{i}$ exhibits no observable microwave-power dependence, and both $Q_\mathrm{i}$ and $\Delta f_\mathrm{r}/f_\mathrm{r}$ persist over a temperature range far above the characteristic scale set by the resonator frequency. Instead, the low-temperature $\Delta f_\mathrm{r}/f_\mathrm{r}$ is consistent with an additional paramagnetic contribution, potentially arising from localized magnetic moments or Andreev bound states in the YBCO film~\cite{Ames_review, ABS_theory}. The low-temperature loss follows an approximately logarithmic temperature dependence and is not captured by existing models of either TLS loss or quasiparticle dissipation in $d$-wave superconductors.


\section{YBCO resonators}
The CPW resonators used in this study are fabricated from YBCO films purchased from Ceraco GmbH. The manufacturer designates two film compositions: S-type films are grown with a slight Cu excess, whereas M-type films contain excess Y. These distinct growth conditions may lead to differences in the microwave properties of the patterned devices. We characterize two devices, with chip 1 patterned from an S-type film and chip 2 from an M-type film. Both as-purchased films consist of a 200-nm Au/210-nm YBCO/10-nm $\mathrm{CeO_2}$/r-plane sapphire stack and have $T_\mathrm{c}\approx87~\mathrm{K}$ and residual resistivity ratios $\sim3$. Here the Au layer is for protecting the YBCO film from moisture during storage and the $\mathrm{CeO_2}$ buffer improves lattice matching while preventing inter-diffusion between sapphire and YBCO~\cite{CeO2, YBCO_diffusion}. The majority of the Au film is removed with standard $\mathrm{KI/I_2}$-based Au etchant, leaving only a few small regions for wire bonding. The YBCO structures are then defined using electron-beam lithography and a wet etch process with $\mathrm{H_3PO_4}$ diluted with de-ionized water. A concentration of $0.5-1\%$ of $\mathrm{H_3PO_4}$ by volume results in an etch rate of $\sim3$~nm/s.

\begin{figure}
\centering
    \includegraphics{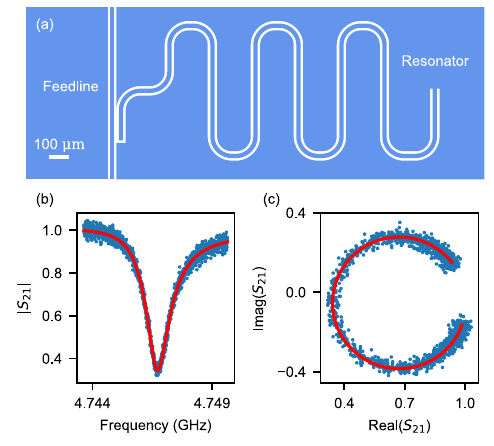}
    \caption{\textbf{Microwave response of a single resonator.} (a) Layout of a single CPW resonator. The microwave signal travels through the feedline capacitively coupled to individual CPW resonators. (b) Typical $|S_{21}|$ vs. frequency of a resonator, measured at $-$130 dBm drive power. The red line shows the best fit to the data. (c) The real vs. the imaginary part of the same $S_{21}$ data. The red line shows the best circle fit to the raw data. }
    \label{fig1}
\end{figure}

\begin{figure}[t]
\centering
    \includegraphics{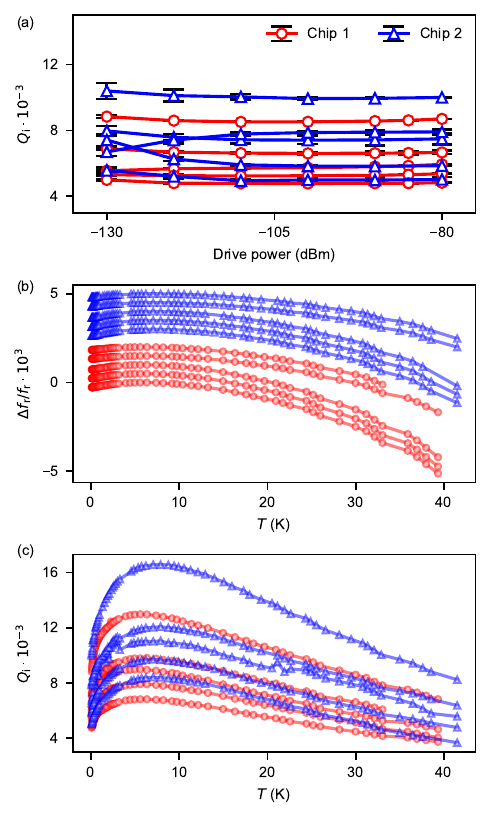}
    \caption{\textbf{Characterization of all the resonators. } (a) Power dependence of $Q_\mathrm{i}$ of all resonators of chip 1 (S-type film) and chip 2 (M-type film). Data points of the same resonator are connected with solid lines. (b) Temperature dependence of $\Delta f_\mathrm{r}/f_\mathrm{r}$ of all resonators measured at -80 dBm. The error bars are smaller than the size of the markers. A temperature-independent offset is added to each resonator for clarity. (c) Temperature dependence of $Q_\mathrm{i}$ of all resonators measured at -80 dBm. }
    \label{fig2}
\end{figure}

The finished device chip contains multiple CPW resonators capacitively coupled to a shared feedline. A representative schematic of a resonator is shown in Fig.~\ref{fig1}(a). These resonators have resonance frequencies $f_\mathrm{r}$ between 4 and 6 GHz. The chip is attached to a Au-plated sample package using GE varnish and is wire bonded to feedlines on a microwave launch board. The package is then attached to an adiabatic demagnetization refrigerator (ADR) with a base temperature of $\sim65~\mathrm{mK}$. The input microwave signal travels through a total of 70 dB of attenuation and an infrared filter before reaching the sample. The output signal passes through two circulators and a high-electron-mobility-transistor (HEMT) amplifier before being further amplified at room temperature. Because the measurements span a wide temperature range, stainless steel cables are used between low-temperature stages instead of NbTi or Nb cables to avoid impedance changes associated with superconducting transitions in the microwave line.

Figure~\ref{fig1}(b, c) shows the transmitted signal near the resonance of a typical resonator at $-$120 dBm of drive power. We extract the intrinsic quality factor $Q_\mathrm{i}$, the coupling quality factor $Q_\mathrm{c}$, and $f_\mathrm{r}$ by fitting the resonance feature to~\cite{DCM, resonatorFit}
\begin{equation}
    S_{21}=p\cdot e^{is-2\pi if\zeta}\left[1-\frac{\left(Q_\mathrm{l}/|Q_\mathrm{c}|\right)\cdot e^{i\phi}}{1+2iQ_\mathrm{l}\cdot\left(f/f_\mathrm{r}-1\right)}\right],
\end{equation}
where $f$ is the frequency being measured, $p$ is the amplitude of the background, $s$ is the phase offset of the background, $\zeta$ is the phase delay of the background, $\phi$ is the phase delay caused by the impedance mismatch between the resonator and the feedline, and $Q_\mathrm{l}=\left(Q_\mathrm{i}^{-1}+\mathrm{Re}\{Q_\mathrm{c}\}^{-1}\right)^{-1}$ is the loaded quality factor. Here $Q_\mathrm{c}$ is a complex number to account for the impedance mismatch to prevent overestimating the extracted $Q_\mathrm{i}$~\cite{DCM}.

When the temperature is fixed at $\sim70~\mathrm{mK}$, the $Q_\mathrm{i}$ of the YBCO resonators on two separate chips are measured to be between $4\cdot10^3$ to $10^4$, consistent with devices fabricated with similar films~\cite{sidharth, YBCO_EPR, german_YBCO_HIM}. As shown in Fig.~\ref{fig2}, $Q_\mathrm{i}$ shows negligible dependence on the microwave drive power up to $-$80 dBm. This indicates that the microwave loss of YBCO samples are likely not dominated by TLS loss in spite of a 10 nm thick $\mathrm{CeO_2}$ buffer layer. Despite a lack of power dependence, both $Q_\mathrm{i}$ and $\Delta f_\mathrm{r}/f_\mathrm{r}$ increase as temperature is increased, qualitatively resembling the behavior caused by TLS defects. At $\sim6~\mathrm{K}$, $Q_\mathrm{i}$ reaches a maximum of $8\times10^3$ to $1.2\times10^4$ among the resonators. As the temperature is further increased, both $Q_\mathrm{i}$ and $\Delta f_\mathrm{r}/f_\mathrm{r}$ continue to decrease, consistent with the effect of quasiparticle loss~\cite{Hirschfeld, hirschfeld2, badass_YBCO_microwave}.

\begin{figure}[t]
\centering
    \includegraphics{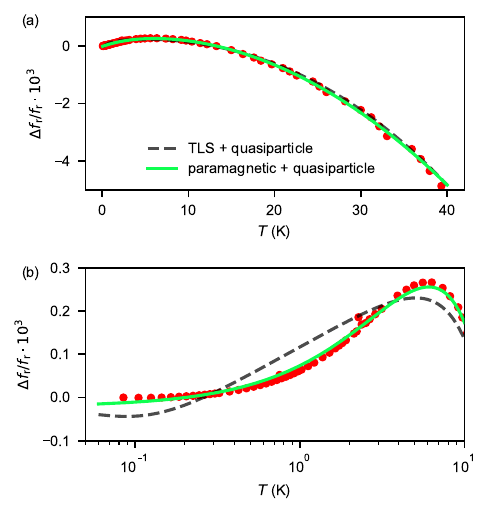}
    \caption{\textbf{Analysis of the temperature dependence of resonator frequency shift.} (a) Temperature dependence of $\Delta f_\mathrm{r}/f_\mathrm{r}$ of a representative resonator fitted to the TLS + quasiparticle model (Eqs.~\ref{eq_df_TLS}, \ref{eq_df_QP}, \ref{eq_Lk_raw}, \ref{eq_lambda_T}) and paramagnetic response + quasiparticle model (Eqs.~\ref{eq_df_QP}, \ref{eq_lambda_T}, \ref{eq_Lk_Curie}). (b) Zoom in plot in the low temperature region revealing the different behavior of the two models. }
    \label{fig3}
\end{figure}

\section{Temperature dependence of frequency shift}

We first analyze the temperature dependence of the $\Delta f_\mathrm{r}/f_\mathrm{r}$ in terms of two standard contributions: the dispersive response of a TLS bath at low temperature and the kinetic-inductance shift associated with thermally generated quasiparticles at higher temperature. 

At low temperatures, $\Delta f_\mathrm{r}/f_\mathrm{r}$ increases with temperature. Such an upturn is often associated with the dispersive response of a TLS bath~\cite{andersonTLS, Phillips_TLS, McRae_resonator}. Although a TLS interpretation is already disfavored by the absence of measurable power dependence shown in Fig.~\ref{fig2}, we nevertheless first test whether the low-temperature frequency shift can be described by the standard TLS expression~\cite{Muller_2019, McRae_resonator},
\begin{equation}
\begin{split}
\left(\frac{\mathrm{d}f_\mathrm{r}}{f_\mathrm{r}}\right)_\mathrm{TLS}
={}& \frac{F\delta_\mathrm{TLS}}{\pi}
\operatorname{Re}\Bigg[
\Psi\left(
\frac{1}{2}
+\frac{i h f_\mathrm{r}}{2\pi k_\mathrm{B}T}
\right) \\
&\qquad
-\ln\left(
\frac{h f_\mathrm{r}}{2\pi k_\mathrm{B}T}
\right)
\Bigg],
\end{split}
\label{eq_df_TLS}
\end{equation}
where $F\delta_\mathrm{TLS}$ is the filling-factor adjusted TLS loss tangent, $\Psi$ is the complex digamma function, $h$ is Planck's constant, and $k_\mathrm{B}$ is the Boltzmann constant.

At high temperatures, $\Delta f_\mathrm{r}/f_\mathrm{r}$ decreases with $T$, consistent with the behavior of thermally generated quasiparticles, which induce a change in the kinetic inductance. We take
\begin{equation}
    \left(\frac{\mathrm{d}f_\mathrm{r}}{f_\mathrm{r}}\right)_\mathrm{qp}=\sqrt\frac{L_\mathrm{g}'+L_\mathrm{k}'\left(0\right)}{L_\mathrm{g}'+L_\mathrm{k}'\left(T\right)}-C,
    \label{eq_df_QP}
\end{equation}
where $L_\mathrm{g}'$ is the geometric inductance per unit length of the CPW, and $L_\mathrm{k}'$ is the kinetic inductance per unit length, and $C\approx1$ is a normalization constant that accounts for a small offset in the reference $f_\mathrm{r}$. For a CPW structure~\cite{Lk_coth_equation, loss_sinh}, 
\begin{equation}
    L_\mathrm{k}'=G\mu_0\lambda_\mathrm{ab}\left(T\right)\coth{\left[\frac{d}{\lambda_\mathrm{ab}\left(T\right)}\right]},
    \label{eq_Lk_raw}
\end{equation}
where $G$ is a geometric factor dependent on the CPW geometry, $\mu_0$ is the vacuum permeability, $\lambda_\mathrm{ab}$ is the in-plane London penetration depth, and $d$ is the film thickness.

For YBCO with finite impurity scattering, the penetration-depth variation has been described by $\lambda_\mathrm{ab}(T)-\lambda_\mathrm{ab}(0)\propto T^2/(T+T^*)$, where $T^*$ characterizes the crossover from impurity-induced $T^2$ behavior at low temperature to approximately linear-$T$ behavior at higher temperature~\cite{Hirschfeld}. A fit to this equation returns $T^*\gg T_\mathrm{c}$, we therefore choose to fit to another widely-used phenomenological model~\cite{YBCO_EPR}
\begin{equation}
    \lambda_\mathrm{ab}\left(T\right)=\lambda_\mathrm{ab}\left(0\right)\cdot\left[1+a\left(\frac{T}{T_\mathrm{c}}\right)^\beta\right],
    \label{eq_lambda_T}
\end{equation}
where the exponent $\beta$ is expected to be around 2.

We fit a representative resonator using the combined TLS + quasiparticle model defined by Eqs.~\ref{eq_df_TLS},~\ref{eq_df_QP},~\ref{eq_Lk_raw}, and~\ref{eq_lambda_T}, as shown in Fig.~\ref{fig3}.
Although the model describes the high-$T$ behavior well, it fails to reproduce the low-$T$ features. Specifically, Eq.~\ref{eq_df_TLS} predicts that, at a characteristic temperature $T\sim hf_\mathrm{r}/2k_\mathrm{B}$, near-resonant TLSs become thermally saturated, producing an initial dip in $\Delta f_\mathrm{r}/f_\mathrm{r}$. For a TLS bath large enough to produce the observed upturn, the accompanying minimum near $100~\mathrm{mK}$ would be clearly resolved in our data. Moreover, the curvature of the measured upturn differs substantially from that predicted for a TLS bath.

We next consider mechanisms intrinsic to YBCO that can produce a low-temperature paramagnetic response and thereby account for the discrepancy with the TLS model. Two natural candidates are defect-induced local magnetic moments~\cite{Ames_review} and Andreev bound states~\cite{ABS_theory, anlage_ABS} near the YBCO surface. Both mechanisms can increase the effective kinetic inductance at low temperature and therefore produce an upturn in the resonance frequency with increasing temperature.

Previous studies have shown that substitution of nonmagnetic ions such as lithium and zinc on Cu sites can induce local magnetic moments that are detectable by nuclear magnetic resonance~\cite{patrickLee2, Cuprate_Zn_NMR, Cuprate_Li_Zn_NMR, Cuprate_Li_Zn_NMR_belowTc}. Structural defects intrinsic to the thin film may similarly disturb the magnetic correlations of the CuO$_2$ planes and generate local moments. These moments produce a Curie-Weiss-like correction to the permeability, $\mu=\mu_0\mu_{\mathrm r}$, $\mu_{\mathrm r}=1+\frac{c}{T+\theta}$, where $c$ determines the strength of the paramagnetic response and $\theta$ is its characteristic temperature scale. Including this correction modifies Eq.~\ref{eq_Lk_raw} to

\begin{equation}
L_{\mathrm{k}}'
=
G\mu_0\sqrt{\mu_{\mathrm r}}
\lambda_{\mathrm{ab}}(T)
\coth\left[
\frac{d\sqrt{\mu_{\mathrm r}}}
{\lambda_{\mathrm{ab}}(T)}
\right].
\label{eq_Lk_Curie}
\end{equation}

A similar low-temperature response can arise from Andreev bound states formed near pair-breaking surfaces or interfaces of YBCO~\cite{ABS_theory, anlage_ABS, anlage_ABS2}. In the clean limit, the paramagnetic response of these states produces an effective penetration-depth correction proportional to $1/T$, with the divergence expected to be cut off at sufficiently low temperature by finite broadening or splitting of the bound states. For small paramagnetic response, localized magnetic moments and Andreev bound states therefore produce corrections proportional to $1/(T+\theta)$ and $1/T$, respectively. The two mechanisms produce qualitatively similar changes in the kinetic inductance and resonance frequency. We therefore do not introduce a separate Andreev bound state fit and instead regard the paramagnetic fit as phenomenologically consistent with contributions from either localized magnetic moments or surface Andreev bound states.

As shown in Fig.~\ref{fig3}, the model defined by Eqs.~\ref{eq_df_QP}, \ref{eq_lambda_T}, \ref{eq_Lk_Curie} captures the measured frequency shift over the full temperature range. The fits yield $\beta=2.07\pm0.06$ for chip 1 and $\beta=2.26\pm0.04$ for chip 2, consistent with the approximately quadratic temperature dependence reported for YBCO thin films~\cite{Hirschfeld, badass_YBCO_microwave, YBCO_EPR}. The extracted zero-temperature penetration depths are $\lambda_{\mathrm{ab}}(0)=232\pm6~\mathrm{nm}$ for chip 1 and $\lambda_{\mathrm{ab}}(0)=160\pm9~\mathrm{nm}$ for chip 2. Although the fit supports a low-temperature paramagnetic contribution beyond the standard TLS response, the present data do not distinguish whether it originates from localized magnetic moments or surface Andreev bound states.

\section{Temperature dependence of microwave loss}

\begin{figure}[t]
\centering
    \includegraphics{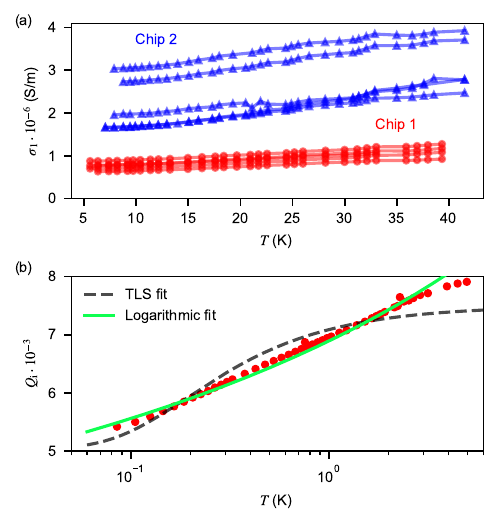}
    \caption{\textbf{Analysis of the temperature dependence of the intrinsic quality factor.} (a) Effective $\sigma_1$ of all resonators in the high temperature regime. (b) Temperature dependence of $Q_\mathrm{i}$ of a representative resonator at low temperature fitted to Eq.~\ref{eq_TLS_loss} and \ref{eq_log_fit}.  }
    \label{fig4}
\end{figure}

Similar to $\Delta f_\mathrm{r}/f_\mathrm{r}$, $Q_\mathrm{i}$ has a temperature dependence that resembles quasiparticle-dominated loss at high $T$ and TLS-dominated loss at low $T$~\cite{andersonTLS, Phillips_TLS, McRae_resonator}. The quasiparticle contribution to $Q_\mathrm{i}$ is~\cite{loss_sinh} 
\begin{equation}
    Q_\mathrm{qp}=\frac{2}{\alpha\mu_0\omega\lambda_\mathrm{ab}^2\sigma_1\left[1+\frac{2d/\lambda_\mathrm{ab}}{\sinh{\left(2d/\lambda_\mathrm{ab}\right)}}\right]},
    \label{eq_loss_qp}
\end{equation}
where $\alpha=L_\mathrm{k}'/\left(L_\mathrm{g}'+L_\mathrm{k}'\right)$ is the kinetic inductance ratio, and $\sigma_1$ is the real part of complex microwave conductivity, which can be described as
\begin{equation}
    \sigma_1=\frac{n_\mathrm{qp}e^2}{m^*}\cdot\frac{\tau}{1+\omega^2\tau^2},
\end{equation}
where $n_\mathrm{qp}$ is the quasiparticle density, $e$ is the electron charge, $m^*$ is the effective mass, $\tau$ is the quasiparticle scattering time, and $\omega$ is the angular frequency. Using $\lambda_\mathrm{ab}\left(T\right)$ and $L_\mathrm{k}'$ extracted from the temperature dependence of $\Delta f_\mathrm{r}/f_\mathrm{r}$ and assuming the high temperature $Q_\mathrm{i}$ is limited by quasiparticles, we calculate $\sigma_1$ above the inflection point of $Q_\mathrm{i}$. As shown in Fig.~\ref{fig4}(a), $\sigma_1$ falls in the range of $10^6~\mathrm{S/m}$ and increases with temperature, consistent with previous observations~\cite{badass_YBCO_microwave, tau1, hirschfeld2}.



Additionally, we attempt to fit the low temperature behavior of $Q_\mathrm{i}$ with the TLS model
\begin{align}
    Q_\mathrm{TLS}&=\frac{\sqrt{1+\frac{\left<n_\mathrm{p}\right>}{n_\mathrm{c}}}}{F\delta_\mathrm{TLS}\cdot\tanh{\left(\frac{hf_\mathrm{r}}{2k_\mathrm{B}T}\right)}}\\
    &=\frac{1}{F\delta_\mathrm{TLS}'\cdot\tanh{\left(\frac{hf_\mathrm{r}}{2k_\mathrm{B}T}\right)}},\label{eq_TLS_loss}
\end{align}
where $Q_\mathrm{TLS}$ is the TLS contribution to $Q_\mathrm{i}$, $\left<n_\mathrm{p}\right>$ is the average photon number inside the resonator, and $n_\mathrm{c}$ is the critical photon number for power saturating the TLS. As only negligible power dependence is observed in Fig.~\ref{fig2}, we do not explicitly fit for $\left<n_\mathrm{p}\right>$ and $n_\mathrm{c}$ and absorb them into $F\delta_\mathrm{TLS}'=\frac{F\delta_\mathrm{TLS}}{\sqrt{1+\left<n_\mathrm{p}\right>/n_\mathrm{c}}}$. 

We show the low temperature $Q_\mathrm{i}$ data of a representative resonator below 6 K and its fit to Eq.~\ref{eq_TLS_loss} in Fig.~\ref{fig4}(b). Although the TLS model does predict an increasing $Q_\mathrm{i}$ with temperature, the resulting $Q_\mathrm{i}$ fit increases and then begins to saturate within the temperature scale of $hf_\mathrm{r}/k_\mathrm{B}$, while the measured $Q_\mathrm{i}$ continues to increase until $\sim6~\mathrm{K}$. Notably, the low temperature $Q_\mathrm{i}$ data can be well captured by a phenomenological model that scales logarithmically with temperature,
\begin{equation}   Q_\mathrm{i}^{LT}=A+B\cdot\ln{\left(\frac{T}{1~\mathrm{K}}\right)}.
\label{eq_log_fit}
\end{equation}

As $d$-wave superconductors may have non-negligible $n_\mathrm{qp}$ at low temperatures~\cite{Hirschfeld}, quasiparticle loss may also dominate the low temperature regime. As neither $\lambda_\mathrm{ab}$ nor $n_\mathrm{qp}$ is expected to have significant temperature dependence at low temperatures, the temperature dependence of $Q_\mathrm{i}$ may originate from changes in the quasiparticle scattering time $\tau$. One possible mechanism that would result in a logarithmic temperature dependence in $\tau$ is Kondo scattering~\cite{kondo}, which has been observed in cuprates with heavy impurities~\cite{Cuprate_Kondo_transport, Cuprate_Kondo_transport_B_field, Cuprate_Kondo_transport_radiation}. However, previous studies have reported $\tau$ in the range of 1 ps for YBCO~\cite{badass_YBCO_microwave, tau1}. This results in $\omega\tau\ll1$, which leads to $\sigma_1\propto\tau$ and $Q_\mathrm{qp}\propto1/\tau$. Typically Kondo scattering would lead to a scattering rate $1/\tau\propto-\mathrm{ln}\left(T\right)$, which is in the opposite direction from the observed trend. We therefore conclude that Kondo scattering is likely not responsible for the observed logarithmic trend.

\section{discussion}

Although the microscopic origin of the logarithmic loss remains unknown, it is plausible that it is related to the same low-temperature degrees of freedom responsible for the paramagnetic frequency shift. Both contributions become prominent over a similar temperature range, and magnetic moments or Andreev-bound-state can both in principle produce a dissipative microwave in addition to the reactive response. If the loss and frequency upturn are connected, identifying the origin of the paramagnetic response becomes an important step toward understanding the unresolved dissipative channel.

In the thin film limit ($d\ll\lambda_{\mathrm{ab}}$), Eq.~\ref{eq_Lk_Curie} reduces to $L_\mathrm{k}'\simeq G\mu_0\frac{\lambda_{\mathrm{ab}}^2}{d}$~\cite{Lk_coth_equation}, which cancels out the $\mu_\mathrm{r}$ dependence. In contrast, the paramagnetic response of surface Andreev bound state enters through an additional surface-current response and will persist in the thin-film limit. Measurements of otherwise comparable resonators fabricated from thinner films may therefore help distinguish the two mechanisms. Magnetic-field and microwave-power dependence could provide additional tests. In particular, the microwave drive used in the present measurements is many orders of magnitude weaker than the power scale reported to suppress the Andreev bound state response~\cite{anlage_ABS2, Ames_review}. Consequently, the absence of measurable power dependence in our data does not exclude an Andreev bound state contribution.

Despite the anomalous low-temperature contribution, the resonators reach maximum internal quality factors of approximately $7\times10^3$ to $1.7\times10^4$ near 6 to $8~\mathrm{K}$. These values establish a quantitative loss benchmark for microwave resonators made from high-$T_\mathrm{c}$ materials. These $Q_\mathrm{i}$ values may enable parametric amplifiers~\cite{KI_TWPA}, microwave detectors~\cite{mKID}, and other superconducting microwave circuits~\cite{cuprateTwistronics} that could benefit from the high $T_\mathrm{c}$, high $H_\mathrm{c}$, and unique $d$-wave behavior of YBCO.

\section{Conclusion}
In summary, we characterize CPW resonators fabricated using YBCO thin films from $\sim70~\mathrm{mK}$ to $\sim40~\mathrm{K}$. The high temperature response is consistent with the temperature dependence of $\lambda_\mathrm{ab}$ and $\tau$ expected from $d$-wave superconductors~\cite{Hirschfeld, hirschfeld2, badass_YBCO_microwave}. Although both $Q_\mathrm{i}$ and $\Delta f_\mathrm{r}/f_\mathrm{r}$ qualitatively agree with the behavior expected from the TLS model, their low-temperature trend extends considerably beyond the temperature scale of $hf_\mathrm{r}/k_\mathrm{B}$ expected from a TLS bath. We show that the low temperature $\Delta f_\mathrm{r}/f_\mathrm{r}$ is consistent with a paramagnetic response from dilute magnetic moments or Andreev bound states in the YBCO film~\cite{Ames_review, ABS_theory} and $Q_\mathrm{i}$ follows an approximate logarithmic temperature dependence with unknown microscopic origin. Our work reveals an unresolved low-temperature loss channel in YBCO and show that microwave response from unconventional superconducting films may mimic signatures commonly associated with TLS.

\begin{acknowledgments} 
The authors would like to thank Steven Anlage for insightful discussion. This work is supported by the Gordon and Betty Moore Foundation, DOI 10.37807/gbmf11557 and the National Science Foundation award No. 2427093. The authors acknowledge the use of facilities at the Institute of Materials Science and Engineering in Washington University. 
\end{acknowledgments}

\appendix 

\setcounter{figure}{0}
\setcounter{table}{0}
\renewcommand{\thefigure}{A\arabic{figure}}
\renewcommand{\thetable}{A\arabic{table}}
\section{Estimation of the penetration depth} \label{app:lambda}

\begin{figure}
\centering
    \includegraphics{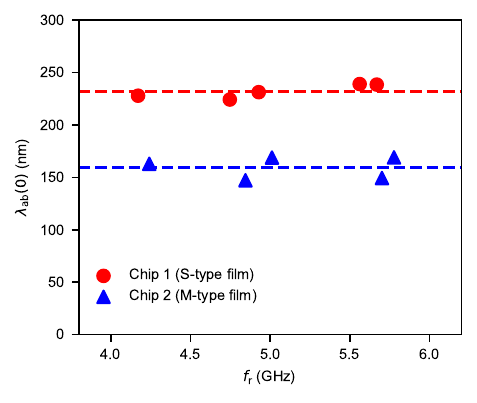}
    \caption{\textbf{Estimated penetration depth of all resonators.} Estimated $\lambda_\mathrm{ab}\left(0\right)$ vs. $f_\mathrm{r}$ of all measured resonators. Within each chip, the extracted values are consistent across different resonator geometries and frequencies, while a systematic difference is observed between the two chips. }
    \label{fig_a1}
\end{figure}

We first extract the expected resonator frequency without any kinetic inductance contribution $f_\mathrm{s}$ by simulating the resonator structures in ANSYS High Frequency Structure Simulator (HFSS) using the actual device dimensions measured with atomic force microscopy and scanning electron microscopy. Then we estimate $L_\mathrm{g}'$ by simulating the cross section of the resonators using ANSYS Q3D. Combined with the measured $f_\mathrm{r}$, $L_\mathrm{k}'$ can be estimated through
\begin{equation}
    L_\mathrm{k}'=L_\mathrm{g}'\cdot\frac{f_\mathrm{s}^2-f_\mathrm{r}^2}{f_\mathrm{r}^2}.
\end{equation}
We then estimate $G$ by sweeping a sheet impedance $\omega L_\mathrm{k}'/G$ on the superconducting film and simulating the resonator frequency.

The extracted $L_\mathrm{k}'$ and $G$ is then used to fit for $\lambda_\mathrm{ab}$. We first set $a=c=0$ to obtain a rough estimate of $\lambda_\mathrm{ab}\left(0\right)$. We then fix $\lambda_\mathrm{ab}\left(0\right)$ and fit $\Delta f_\mathrm{r}/f_\mathrm{r}$ to obtain $a$, $c$, $\theta$, and $n$. We then fit these parameters to extract an updated $\lambda_\mathrm{ab}\left(0\right)$ based on $f_\mathrm{r}$ measured at $\sim70~\mathrm{mK}$. This process is repeated until $\lambda_\mathrm{ab}\left(0\right)$ does not change more than $0.1\%$ between iterations. The resonators on each chip fall into two designs. Both designs have $\sim18~\mathrm{\mu m }$ trace width but they have a different gap width of either $\sim14~\mathrm{\mu m}$ or $\sim50~\mathrm{\mu m}$. As shown in Fig.~\ref{fig_a1}, despite the different geometries, our approach is able to find a global $\lambda_\mathrm{ab}\left(0\right)$ across each chip which is dominated by the growth method of the respective film.

\section{Fitting results for all resonators}
In Tab.~\ref{tab:a1}, we summarize the basic parameters of all the resonators measured in this study, fitting parameters obtained by fitting the temperature-dependent $\Delta f_\mathrm{r}/f_\mathrm{r}$ data to Eqs.~\ref{eq_df_QP}, \ref{eq_lambda_T}, \ref{eq_Lk_Curie}, and fitting parameters obtained by fitting the low temperature $Q_\mathrm{i}$ data to the phenomenological Eq.~\ref{eq_log_fit}.

\begin{widetext}

\refstepcounter{table}
\begin{center}
\textbf{TABLE~\thetable. Resonator parameters and fit results.}
\label{tab:a1}


\scriptsize
\setlength{\tabcolsep}{4pt}
\begin{ruledtabular}
\begin{tabular}{ccccc cccc cc}

Chip &
Res. &
$f_\mathrm{r}$ &
$Q_\mathrm{i,max}$ &
$Q_\mathrm{i,base}$ &
\multicolumn{4}{c}{$\Delta f_\mathrm{r}/f_\mathrm{r}$ fit}  &
\multicolumn{2}{c}{$Q_\mathrm{i}$ fit} \\
\cline{6-9}
\cline{10-11}

 &  & (GHz) &  &  &
$a$ &
$\beta$ &
$c$ (K)&
$\theta$ (K)&
$A\cdot10^{-3}$ &
$B\cdot10^{-2}$ \\
\hline

1 & 1A & 4.1712 & $8986\pm98$ & $6003\pm14$ & $0.38\pm0.01$ & $2.09\pm0.06$ & $0.85\pm0.35$ & $8.28\pm2.02$ & $7.974\pm0.09$ & $7.92\pm0.08$ \\
1 & 1B & 4.7466 & $12974\pm6$ & $8809\pm4$ & $0.44\pm0.01$ & $2.12\pm0.06$ & $0.86\pm0.32$ & $8.39\pm1.85$ & $11.370\pm0.006$ & $10.74\pm0.05$ \\
1 & 1C & 4.9279 & $7919\pm22$ & $5423\pm10$ & $0.38\pm0.01$ & $2.10\pm0.06$ & $0.86\pm0.36$ & $8.39\pm2.11$ & $6.963\pm0.004$ & $6.51\pm0.03$ \\
1 & 1D & 5.5611 & $9817\pm72$ & $7060\pm34$ & $0.35\pm0.02$ & $1.96\pm0.07$ & $1.55\pm0.70$ & $10.81\pm2.65$ & $8.700\pm0.010$ & $6.89\pm0.10$ \\
1 & 1E & 5.6683 & $6831\pm13$ & $4765\pm5$ & $0.37\pm0.01$ & $2.11\pm0.06$ & $0.82\pm0.33$ & $8.13\pm1.97$ & $5.976\pm0.005$ & $5.25\pm0.05$ \\

\hline

2 & 2A & 4.2414 & $11052\pm152$ & $6537\pm57$ & $0.55\pm0.01$ & $2.20\pm0.04$ & $0.63\pm0.12$ & $6.22\pm0.73$ & $9.25\pm0.05$ & $11.4\pm0.4$ \\
2 & 2B & 4.8449 & $16564\pm84$ & $9996\pm32$ & $0.67\pm0.02$ & $2.25\pm0.04$ & $0.56\pm0.10$ & $5.76\pm0.65$ & $13.29\pm0.07$ & $15.0\pm0.5$ \\
2 & 2C & 5.0111 & $9698\pm12$ & $5834\pm5$ & $0.53\pm0.01$ & $2.24\pm0.04$ & $0.56\pm0.10$ & $5.92\pm0.71$ & $7.74\pm0.05$ & $8.9\pm0.3$ \\
2 & 2D & 5.7008 & $12074\pm237$ & $7838\pm92$ & $0.68\pm0.02$ & $2.31\pm0.04$ & $0.46\pm0.09$ & $5.35\pm0.66$ & $9.90\pm0.05$ & $9.4\pm0.3$ \\
2 & 2E & 5.7764 & $8415\pm159$ & $5019\pm73$ & $0.54\pm0.16$ & $2.32\pm0.05$ & $0.45\pm0.09$ & $5.32\pm0.73$ & $6.64\pm0.04$ & $7.7\pm0.3$ \\

\end{tabular}
\end{ruledtabular}
\end{center}

\end{widetext}

%

\end{document}